\documentclass[12pt,twoside]{article}
\usepackage{graphicx}
\usepackage{amsmath}
\usepackage{latexsym}

\newcommand{\ket}[1]{ \left | #1 \right \rangle}

\title{Transmitting-state invertible cellular automata}
\author{Benjamin Schumacher\footnote{Department of Physics, Kenyon College.  Email schumacherb@kenyon.edu} 
    \,\, and Michael D. Westmoreland\footnote{Department of Mathematical Sciences, Denison University.  Email westmoreland@denison.edu}}

\begin{document}

\maketitle

\thispagestyle{empty}

\begin{abstract}
  Invertible cellular automata are useful as models of physical 
	systems with microscopically revesible dyanmics.  There are several
	well-understood ways to construct them:  partitioning rules, second-order
	rules, and alternating-grid rules.  We present another way (a 
	generalization of the alternating-grid approach), based on the idea
	that a cell may either transmit information to its neighbors or
	receive information from its neighbors, but not both at the same
	time.  We also examine an interesting simple example of this class
	of rules, one with an additive conserved ``energy''.
\end{abstract}

\section{Background}

In cellular automata, the discrete states of the cells in a 
regular grid are updated in time according to a local rule---that
is, the new state of a cell depends only on its old state and the old
state of a finite neighborhood around it.  The same rule applies to all
cells.  The locality and uniformity of this procedure give 
CA's a ``physics-like'' quality, which is why CA's have attracted 
much attention as simplified discrete simulations of physical 
systems---or in a more speculative vein, as
models of the fundamental physical laws themselves \cite{margolusphd,wolframbook}.

A cellular automaton rule is said to be {\em invertible} 
(or {\em reversible}) if it maps
configurations of a (finite or infinite) grid to 
new configurations in a 1-1 way \cite{toffolimargolus}.  
The earlier state of the 
grid can be uniquely determined from the later state.
Invertible CA's thus capture a further principle 
of physics:  the microscopic reversibility of the 
dynamical laws of an isolated system.
Invertible CA models have been used to study reversible computation
and to simulate hydrodynamic and magnetic systems \cite{hydroandmagnetic}.

The definition of CA invertibility is a global one, but a theorem 
of Richardson \cite{richardson} guarantees that it has a local meaning.
If a CA rule $\mathcal{U}$ yields an invertible map 
on global grid configurations, 
then the inverse map is always given by another CA rule $\mathcal{U}'$, 
possibly with a much larger neighborhood than $\mathcal{U}$.
It can be quite difficult to tell whether a given CA rule
is invertible---that is, whether any such $\mathcal{U}'$ exists.  
In fact, in grids of more than one dimension, 
there is no effective computational procedure for deciding whether
a rule is invertible in every case \cite{toffolimargolus}.
From the model-builder's perspective, however, it
is more important to devise techniques for {\em constructing} 
CA's that are guaranteed to be invertible.

We have identified three general approaches that are used
to construct invertible CA rules.  
\begin{itemize}
\item
The first is the {\em partitioning rule}
approach.  The grid is partitioned uniformly into finite blocks, and
the configuration of each block is updated according to some map that is
both 1-1 and onto.
Since there are only a finite number of states for each block, it
is easy to confirm that a given update map is invertible.  At the next
step, the block partition is shifted according to a predetermined rule,
and the process is repeated.

To visualize this, consider a 1-D grid of cells, each of which has $n$ possible
states.  We partition the grid into blocks of length 2, so that each block
contains an even-odd pair of cells ($\cdots\,|\,4,5\,|\,6,7\,|\,8,9\,|\cdots$).  Each
2-cell block is updated according to some 1-1 and onto map $U$ of the $n^2$ 
possible block states.  In the next time step, we partition the grid into
odd-even blocks ($\cdots, 4\,|\,5,6\,|\,7,8\,|\,9,\cdots$) and apply the same update rule
$u$ to these.  Further steps repeat this alternation, as shown in
Figure~\ref{fig-partitioning}.
\begin{figure}
\begin{center}
  \includegraphics[height=2.0in]{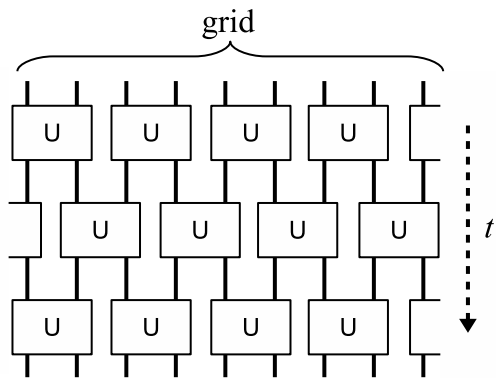}
\end{center}
\caption{Structure of a partitioning rule in 1-D.\label{fig-partitioning}}
\end{figure}

This is plainly an invertible CA, since the inverse block map $U^{-1}$ 
exists and we can apply it to the reverse sequence of partitions.

\item
The second method of constructing invertible CA rules is the {\em second-order
rule} approach.  This is easiest to understand in the context of a binary
CA with cell states 0 and 1 (though it can be generalized to any number 
of cell states).  At any given time $t$, we consider 
the present state $c^{t}$ of a cell, the
states $n_{1}^{t}, n_{2}^{t}, \ldots$ of its neighbors, and 
the previous state $c^{t-1}$ of the cell.  The rule is determined by
choosing an arbitrary binary function $f(c,n_1,n_2,\ldots)$ of the cell
and its neighbors.  The update rule is then
\begin{equation}
	c^{t+1} = c^{t-1} \oplus f(c^{t},n_{2}^{t},n_{1}^{t},\ldots) ,
\end{equation}
where $\oplus$ represents addition modulo 2.

This is called a ``second-order'' rule because the new cell state depends on
cell states of two successive times.  Any such rule is
invertible---indeed, its inverse is the same rule provided
the time steps are reversed.  That is,
\begin{equation}
	c^{t-1} = c^{t+1} \oplus f(c^{t},n_{2}^{t},n_{1}^{t},\ldots) ,
\end{equation}
as is easily seen.

\item
The third way to build an invertible CA is to use use the {\em alternating-subgrid}
approach.  Consider a 2-D square grid of cells, and imagine that the cells are
alternately colored black and white in the manner of a checkerboard.  (This coloring 
is a kind of background {\em texture} that exists independently of the cell states.)
The neighborhood of a cell contains only itself and the cells immediately adjacent to
it.  Thus, the neighborhood of a black cells consists of itself and four white neighbors,
and {\em vice versa}.

The trick is now to update black and white cells in alternate time steps.  If $t$ is
even only the black cells are updated, and if $t$ is odd only the white cells are
updated.  Thus, no cell is updated in the same step as its neighbors.  If the cell
updates according to
\begin{equation}
	c^{t+1} = U(c^{t}, n^{t}, e^{t}, s^{t}, w^{t})
\end{equation}
we know that the the states $(n,e,s,w)$ of the four neighbors (north, east, south, west)
are not updated at the same time.

This by itself is not enough to guarantee invertiblity.  To do this, we need to
require that, for any neighbor configuration $(n,e,s,w)$, the function $U$
is 1-1 and onto on the set of states $c$.  Since the states are drawn from a finite
set, it suffices to require that $c_1 \neq c_2$ implies
\begin{equation}
	U(c_1,n,e,s,w) \neq U(c_2,n,e,s,w)
\end{equation}
for any given choice of $(n,e,s,w)$.  Given this condition, there is another function
$U'$ of the neighborhood that reverses the action of $U$ on each updating cell.  The
inverse of the alternating-subgrid CA is just another alternating-subgrid CA using
$U'$ instead of $U$.
\end{itemize}

The problem of constructing invertible CA rules is the problem of conserving information.
The value of the cell state $c^{t}$ at time $t$ constitutes information, and it
must always be possible to recover this information from the state of the grid at the
next time $t+1$.  This might be very complicated.  The subsequent state $c^{t+1}$ of the 
same cell generally depends on the states of several cells (the whole neighborhood) 
at time $t$, so it is usually not possible to reconstruct $c^{t}$ from $c^{t+1}$ alone.
The information about $c^{t}$ is held jointly by many cells at the later time.  But
all of those cells have meanwhile been influenced by many other cells.  To construct
an invertible rule, we must somehow devise this local network of information exchange 
to globally preserve all of the cell state information over time.

\begin{figure}
\begin{center}
  \includegraphics[height=1.5in]{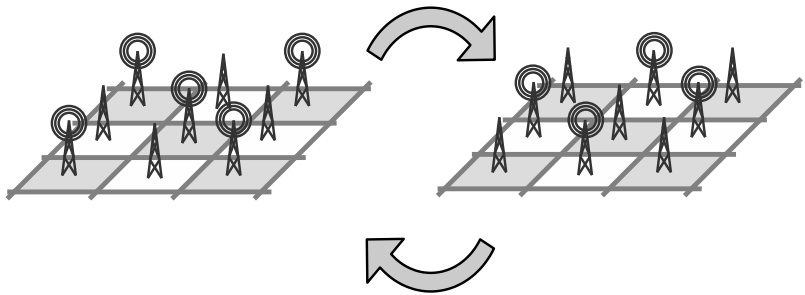}
\end{center}
\caption{In an alternating-subgrid CA, cells are either transmitting information
  about their own state or receiving information from transmitting neighbors.
	\label{fig-alternating}}
\end{figure}

The alternating-subgrid technique described above addresses this challenge in a
very simple way.  At any given step, no cell is both updating its own state and
influencing the states of its neighbors.  That is, the cell is either
listening to its neighbors (and updating its state accordingly) or talking to
its neighbors (and influencing their states), but never both.  Since the neighbors 
that influence a cell's new state retain their own states (for that time step), they 
retain the necessary ``context'' information to reverse the cell update 
operation.  See Figure~\ref{fig-alternating}.

Our approach is to generalize this basic idea, taking what we may call a
{\em transmitting-state} appraoch.  At any given time $t$, each particular
cell may be in a ``transmitting'' or ``non-transmitting'' state.  There may 
be several distinct states of each type.  The update rule for a transmitting 
state does not depend at all on the cell's neighbors.  The update rule for a
non-transmitting state only depends on the cell's own value and that of its
transmitting neighbors.  That is, cells in transmitting states are not 
subject to outside influences, and those that are not transmitting are
only influenced by transmitting neighbors.

In the sections that follow, we make this intuitive idea rigorous and show
how to use it to create an invertible cellular automaton rule.  We briefly 
explore an example of a 3-state/1-D invertible rule of this type, and show
that this rule conserves an additive ``energy'' quantity on the grid.  (Such
conservative rules are useful for exploring thermodynamics in discrete systems.)
Finally, we add some remarks and pose some unresolved questions.

\section{Transmitting and non-transmitting states}

Our cellular automata will exist on a regular grid in any number
of dimensions.  Each cell has a finite set $S$ of states, and 
its neighborhood consists of itself and $N$ neighboring cells.
The update rule is a function of the form $U(c,s_{1},\ldots,s_{N})$,
where $c$ is the cell state and $s_{k}$ is the state of the $kth$
neighbor cell.  That is,
\begin{equation}
	c^{t+1} = U(c^{t},\ldots,s_{k}^{t},\ldots) .
\end{equation} 
Now we identify a subset $T \subseteq S$ of ``transmitting''
states; non-transmitting states are those not in $T$.  There
is also a subset $T^{+} \subseteq S$ of the same size as $T$,
which we call ``post-transmitting'' states.  We now require
five things about the update rule $U$.
\begin{enumerate}
    \item  $c \in T$ iff $U(c,\ldots) \in T^{+}$ for any 
        configuration of neighbor states.  Transmitting states, 
        and only those, evolve into post-transmitting states.
    \item  If $c \in T$ then for any two neighbor states
			$s_{k}, s_{k}' \in S$, $U(c,\ldots,s_{k},\ldots) =
        U(c,\ldots,s_{k}',\ldots)$.
        In other words, if $c$ is a transmitting state, the new
        cell state does not depend on the particular neighbor states.
    \item  If $c,c' \in T$ and $c \neq c'$, then 
        $U(c,\ldots) \neq U(c',\ldots)$.  Two distinct
        transmitting states evolve into two distinct
        post-transmitting states.  The update rule on 
				transmitting states (which is effectively a simple 
				map from $T$ to $T^{+}$) is 1-1 (and thus onto).
    \item  If $c \notin T$, then for any $s_{k},s_{k}' \notin T$
        we have $U(c,\ldots,s_{k},\ldots) = U(c,\ldots,s_{k}',\ldots)$.
        The values of non-transmitting neighbor states do not affect
        the update of a non-transmitting cell state.
    \item  If $c,c' \notin T$ and $c \neq c'$, then for a given
			configuration of neighbor states, 
        $U(c,\ldots) \neq U(c',\ldots)$.  (And from the previous 
				requirement, we know that only the transmitting states in
				the neighborhood configuration make any difference.)
\end{enumerate}

Our basic claim is that any such rule is reversible.  We prove this by
explicitly constructing an inverse CA rule $U'$ from the forward
rule $U$.  The inverse rule $U'$ has the same neighborhood configuration
as $U$.  We define the inverse rule as follows.

First, suppose $c' \in T^{+}$.  Then by conditions 1--3 above, there exists
a unique $c \in T$ such that $c' = U(c,\ldots)$.  For any 
neighbor configuration, define $U'(c',\ldots) = c$.

Now suppose $c' \notin T^{+}$.  Each of the neighbor states $s_{k}'$ is 
either in $T^{+}$ or not.  If not, choose $s_{k}$ for that cell to be any
state not in $T$ (since its exact value, by condition 4, will not
matter).  If $s_{k}' \in T^{+}$, then choose $s_{k} \in T$ to be the 
state that maps to $s_{k}'$ under $U$ (and thus that $s_{k}'$ maps to 
$s_{k}$ under $U'$ as already defined).
This means that we can reconstruct the ``context'' of transmitting
neighbor states $s_{k}$ which governed the update of the central cell 
under $U$.  By condition 5, there is a unique $c \in T$ such that 
\begin{equation}
	c' = U(c,\ldots,s_{k},\ldots).
\end{equation}
Define $U'(c',\ldots,s_{k}',\ldots) = c$.

We have thus defined the rule $U'$ so that it exactly inverts the action 
of update rule $U$, both for cells that have states in $T$ at time $t$ 
(that is, $T^{+}$ at time $t+1$) and cells that do not.  
The CA rule $U$ is therefore invertible.

In an alternating-grid rule, there is a fixed alternating pattern
of transmitting and non-transmitting cells, the checkerboard ``texture''
of black and white cells.  In a transmitting-state
rule, the pattern of transmitting states is itself dynamically 
determined by the action of the CA rule $U$.  This allows a far greater
range of possible rules and behaviors.

\section{An example}

So far, we have given general conditions for transmitting-state rules
without a recipe for creating them.  Now we give one such recipe,
though the transmitting-state class includes many other kinds of
examples.  We take the set of cell states to be $S=\{ 0, 1, \ldots , n \}$,
containing $n+1$ elements.  The set $T$ of transmitting states
contains the $n-1$ elements $\{1, \ldots , n-1\}$, and 
$T^{+} = \{ 2, \ldots , n \}$.  The non-transmitting states
only include states 0 and $n$.

The update rule for a transmitting state simply increments the
state by 1.  That is, for $c \in T$, $U(c,\ldots) = c+1$.  The
transmitting states thus advance regularly via
$1 \rightarrow \ldots \rightarrow n-1 \rightarrow n$.  This
automatically satisfies conditions 1--3 for a transmitting-state
rule.

How is a cell in a non-transmitting state updated?  Non-transmitting
states 0 or $n$ can result in non-post-transmitting states 0 or 1.
For $c=0$, we choose $U(0,\ldots)$ to be any $\{0,1\}$-valued function 
of the transmitting states and their arrangement in the neighborhood.
(Since the $U$ value depends only on the neighbor states that are 
in $T$, it satisfies condition 4 above.)
We then require $U(n,\ldots) = \overline{U(0,\ldots)}$, the binary
complement of the value for the same arrangements of neighborhood
transmitting states.  This guarantees that we will also satisfy 
condition 5, yielding a transmitting-state invertible CA rule.

As a specific example, consider a rule on a 1-D grid with a 3-cell
neighborhood, so the update rule is of the form $U(l,c,r)$, a 
function of the left, center and right cells of the neighborhood.
The set of states $S = \{0,1,2\}$.  The only transmitting state
is 1 and the only post-transmitting state is 2.  Our update rule
is given by
\begin{eqnarray*}
    l1r & \longrightarrow & 2 \mbox{ for any $l,r$} \\
    l0r & \longrightarrow & \left \{ 
        \begin{array}{cl} 1 & \mbox{$l=1$ or $r=1$ but not both}\\
                          0 & \mbox{otherwise}
        \end{array} \right . \\        
    l2r & \longrightarrow & \left \{ 
        \begin{array}{cl} 0 & \mbox{$l=1$ or $r=1$ but not both}\\
                          1 & \mbox{otherwise}
        \end{array} \right .
\end{eqnarray*}
This rule happens to be the 3-state invertible CA numbered 
1123289366095 in Wolfram's enumeration of such rules (as shown 
in \cite{wolframbook}, p. 436).

\begin{figure}
\begin{center}
	\includegraphics[height=2.5in]{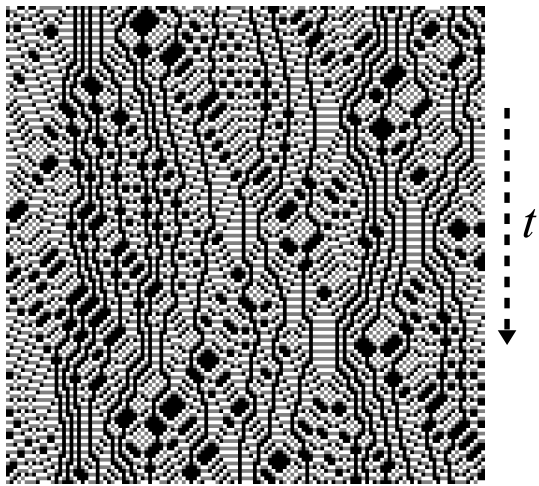}
\end{center}\caption{Spacetime diagram of the behavior of a 
  simple transmitting-state CA in 1-D.  Black and white
	cells are in non-transmitting states, while gray cells 
	represent transmitting states.
	\label{fig-spacetime}}
\end{figure}

The evolution of a random initial grid under this CA rule 
is shown in Figure~\ref{fig-spacetime}.
As can be seen, this CA rule allows the existence of both
persistent spatial structures and long-range information
exchange.

The inverse rule for this example is easily found.  We can write
it this way:
\begin{eqnarray*}
    l2r & \longrightarrow & 1 \mbox{ for any $l,r$} \\
    l0r & \longrightarrow & \left \{ 
        \begin{array}{cl} 1 & \mbox{$l=2$ or $r=2$ but not both}\\
                          0 & \mbox{otherwise}
        \end{array} \right . \\        
    l1r & \longrightarrow & \left \{ 
        \begin{array}{cl} 0 & \mbox{$l=2$ or $r=2$ but not both}\\
                          1 & \mbox{otherwise}
        \end{array} \right .
\end{eqnarray*}
In fact, this is exactly the same rule, with the roles of 
state 1 and state 2 exchanged.

\section{A conserved ``energy''}

Many invertible CA rules, especially those of interest for modeling
physical systems, have a conserved quantity, which we may conveniently
call ``energy''.  The energy is an additive local function.  That
is, for each small neighborhood $n$ (not necessarily identical to a 
neighborhood for the CA rule) there is a local energy $\varepsilon_n$,
and the total energy is just the sum of the local energies of
all the overlapping neighborhoods in the grid.

The 3-state rule we describe above turns out to have a straightforward
conserved energy.  The local energy $\varepsilon$ depends on the joint state
of a pair adjacent cells:
\begin{equation}
  \begin{array}{r}
	\varepsilon(0,0) = \varepsilon(1,2) = \varepsilon(2,1) = 0 \\
	\varepsilon(0,1) = \varepsilon(1,0) = \varepsilon(0,2) = \varepsilon(2,0) = 1 \\
	\varepsilon(1,1) = \varepsilon(2,2) = 2
	\end{array} 
\end{equation}
The local energy $\varepsilon$ resides in the ``bond'' between adjacent cells.  
The total energy $E$ is the sum of all the bond energies.

Energy $E$ will be conserved provided there exists a local ``energy flow'' 
function $f$, which is defined at each cell and depends 
on the states of that cell and its neighbors.  
The left-right symmetry of both the update rule and
the local energy function means that $f$ needs to be an antisymmetric function.
That is, given cell states $l$, $c$ and $r$ at time $t$
(determining the cell state $c'$ at time $t+1$), the flow $f$ satisfies
\begin{equation}
	f(l,c,r) = -f(r,c,l).
\end{equation}
The flow $f$ is to be interpreted as the rightward flow of energy,
so that a negative value of $f$ is a flow in the opposite direction.

Just as $\varepsilon$ is localized between cells, we can regard
$f$ to be localized between time steps $t$ and $t+1$.
\begin{figure}
\begin{center}
  \includegraphics[height=1.25in]{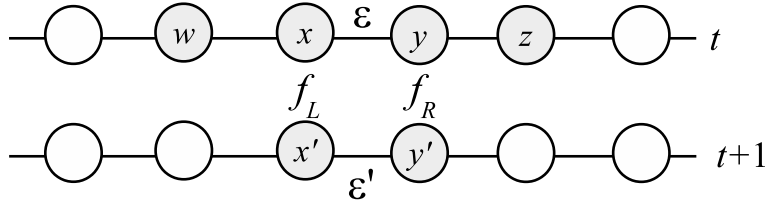}
\end{center}
\caption{Local energy $\varepsilon$ and energy flow $f$ in a small section
  of a spacetime diagram for our example CA rule.\label{fig-energyflow}}
\end{figure}
Consider then the situation shown in Figure~\ref{fig-energyflow}.  At time
$t$, the pair of cells with states $x$ and $y$ have a local bond energy 
$\varepsilon = \varepsilon(x,y)$.  These cell states update to $x'$ and $y'$ 
at $t+1$, yielding a new bond energy $\varepsilon' = \varepsilon(x',y')$.
(The updated cell states also depend on the adjacent states $w$ and $z$ at time $t$.)

The two cells are associated with energy flows $f_L = f(w,x,y)$ and $f_R = f(x,y,z)$,
also shown in Figure~\ref{fig-energyflow}.  Energy is conserved provided we can
show that the change $\Delta \varepsilon$ in local energy is entirely due to
the net inward flow $f$ into that region during the cell update---that is,
$\varepsilon' - \varepsilon = f_{L} - f_{R}$.  We must show that, for all 
cell states $w$, $x$, $y$ and $z$, with $x'$ and $y'$ given by the CA update 
rule,
\begin{equation}
	\varepsilon(x',y') - \varepsilon(x,y) = f(w,x,y) - f(x,y,z) .  \label{eq-localconservation}
\end{equation}
A suitable definition of $f$ for our example can be given as follows:
$f(l,c,r) = 0$ for any set of states with $l=r$ (a symmetric neighborhood state).
Also $f(l,c,r) = 0$ for all other states, with the following exceptions:
\begin{equation}
	\begin{array}{ll}
	f(0,0,1) = -1 & f(1,0,0) = +1 \\
	f(1,2,0) = -1 & f(0,2,1) = +1 \\
	f(2,0,1) = -1 & f(1,0,2) = +1 \\
	f(1,2,2) = -1 & f(2,2,1) = +1 
	\end{array}
\end{equation}

We can now in principle check all $3^4 = 81$ possible cases for Equation~\ref{eq-localconservation}.
In practice, we only need to examine a much smaller number of distinct, non-trivial cases.
For instance, suppose $(w,x,y,z) = (1,1,2,2)$, so that $(x',y') = (2,0)$ according to our rule.
The initial and final energies are $\varepsilon = \varepsilon(1,2) = 0$, while $\varepsilon' 
= \varepsilon(2,0) = 1$.  The two flows are $f_L = f(1,1,2) = 0$ and $f_R = f(1,2,2) = -1$.
Equation~\ref{eq-localconservation} clearly holds; and by symmetry it also holds for
the reflected case $(w,x,y,z) = (2,2,1,1)$.

\begin{figure}
\begin{center}
  \includegraphics[height=2.5in]{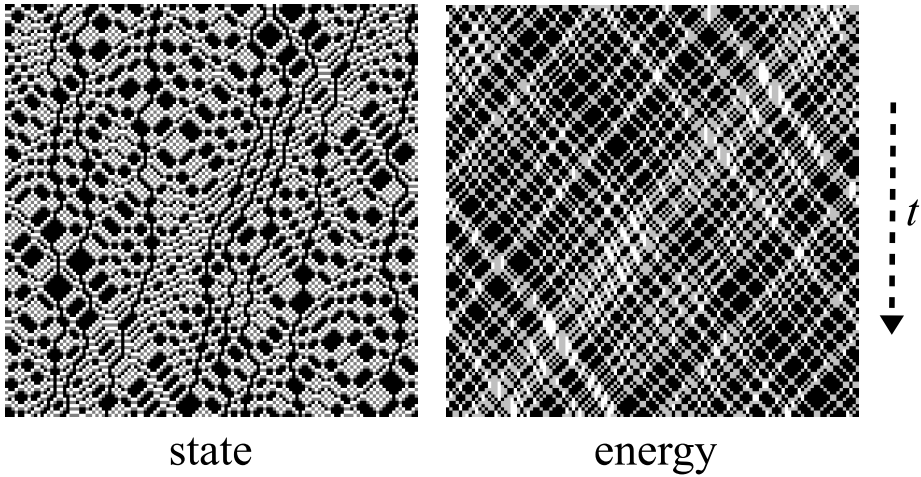}
\end{center}
\caption{The state evolution of our example CA (left), together with the local
  energy $\varepsilon$ (right) for the same initial grid.\label{fig-stateenergy}}
\end{figure}

Figure~\ref{fig-stateenergy} shows both the state evolution and the distribution of 
energy for the same initial grid.  Energy flows are easily visible as 
leftward and rightward movements of elementary energy units.

\section{Remarks and questions}

Transmitting-state cellular automata are a generalization of the alternating-subgrid CA's,
and like them can be used to construct invertible CA rules that produce complex
behavior.  We have given a general definition of transmitting-state rules and
explored a specific example in some detail.  Many questions remain.

The alternating-subgrid class includes rules of special interest for modeling physical systems.  
Consider, for example, the 2-D ``reversible Ising'' CA rule.  In this rule, spins (with $s = \pm 1$)
are located on a checkerboard grid and given an Ising-type energy function 
\begin{equation}
	E = - \sum s_{i} s_{j} ,
\end{equation}
where the sum ranges over all pairs of immediate neighbors $(i,j)$.  At any even time $t$,
each spin on a white square is inverted provided this does not change the energy; 
at any odd time $t$ the same procedure is followed for the spins on black squares.  
The rule is obviously invertible and conserves the Ising energy.  
Thus it is an interesting model of the microscopic
dynamics of a ferromagnetic system.  Does the larger transmitting-state class of CA rules, which
in effect have shifting, dynamically produced ``alternating subgrids'', include
other models for more disordered magnetic systems---e.g., for spin glasses?

We also note that our very simple example of a transmitting-state rule (one of the simplest
non-trivial rules possible) has a simple conserved energy function.  Are such functions 
more common and/or easier to identify in transmitting-state rules?

Reversible cellular automata have been generalized to quantum cellular automata \cite{quantumca}.
Any reversible state function $\mathcal{S} \rightarrow f(\mathcal{S})$ can be made into a
unitary map by applying it to a basis of states:  $\ket{\mathcal{S}} \rightarrow U \ket{\mathcal{S}}
= \ket{f(\mathcal{S})}$.  We can always do this for the global state of a CA grid.
However, locality of the reversible classical rule does not 
guarantee locality of the corresponding quantum rule.  (This objection does not arise for the 
partitioning type of reversible CA, so these can always be generalized to a quantum version.)
The transmitting-state idea is based on the notion of one-way information
flow, and we know that this is impossible in unitary quantum interactions \cite{nooneway}.
How does this lead to a failure of locality in the ``unitarized'' version?  Or are there
examples that can be made into unitary QCA's, despite this difficulty?

We would like to express their gratitude for many helpful conversations 
with Charles Bennett and Tommaso Toffoli.  We also acknowledge the support 
of the Foundational Questions Institute (FQXi), via grant FQXi-RFP-1517.

\end{document}